\documentclass[aps,pre,onecolumn,superscriptaddress,showpacs,amsmath,amssymb]{revtex4-2}

\usepackage[utf8]{inputenc}
\usepackage{amsmath,amsthm}
\usepackage{graphicx}
\usepackage[colorlinks = true,
            linkcolor = blue,
            urlcolor  = blue,
            citecolor = blue,
            anchorcolor = blue]{hyperref}
\usepackage{braket}
\usepackage{xcolor}
\newcommand{\bea}{\begin{eqnarray}}
\newcommand{\eea}{\end{eqnarray}}

\global\long\def\ga{\gamma}

\global\long\def\ell#1{\theta_{#1}}

\global\long\def\al{\alpha}

\global\long\def\ga{\gamma}

\global\long\def\no{\nonumber}

\theoremstyle{thm@}

\theoremstyle{remark}

\begin{document}

\title{
Multilane Asymmetric Exclusion Process with  stationary Bernoulli measure
}


\author{Vladislav Popkov }
\affiliation{Faculty of Mathematics and Physics, University of Ljubljana, Jadranska 19, SI-1000 Ljubljana, Slovenia}
 \affiliation{Department of Physics,
  University of Wuppertal, Gaussstra\ss e 20, 42119 Wuppertal,
  Germany}

\begin{abstract}
We consider an  Asymmetric Exclusion Process evolving on parallel mutually interacting lanes with neighbouring nearest hoppings of hardcore particles.  
Number of particles on each lane is conserved.  We find a choice of the hopping rates,  for which the process has Bernouilli stationary product measure,  and calculate the 
 stationary particle currents as a function of average particle densities.  
\end{abstract}
\maketitle

\textit{Introduction.~~}
 Asymmetric Exclusion process \cite{Pipkin} (ASEP) in one dimension definitely  plays a special role among other many-body non-equilibrium statistical models.  
From the applications perspective,  it is a fundamental minimal model of a traffic flow,   and it serves as a platform for numerous generalizations, 
including those used to monitor real traffic  in cities and on highways with on- and off-ramps. 
From a mathematical perspective,  ASEP is an extremely rich model featuring exciting and remarkable properties: Kardar-Parisi-Zhang universality, 
connections to random matrix theory,   etc. \cite{2006MallickASEP,2011MallickASEP,2000Johansson}.  
The Markov process generator of ASEP can be viewed as a nonhermitian version of a paradigmatic Heisenberg spin $\frac12$ chain Hamiltonian \cite{2001GunterReview}.     
All this richness appears despite the fact that  ASEP,  on an infinite 
lattice,  has an extremely simple stationary state,  namely the product measure,  or Bernoulli,  stationary state,  with zero correlation length. 

 Among numerous ASEP generalizations there is a multilane ASEP \cite{2004MultilanePopkovSalerno},  which turned useful for demonstrating universality classes 
appearing in systems with several conservation laws (the so-called Fibonacci universality,  
characterized by space-time correlations 
with dynamical exponents given by ratio of nearest Fibonacci numbers \cite{2015Fibonacci}),  and a two- and multi-lane ASEPs which manifested
universality in stochastic models with  degenerate characteristic velocities  \cite{2024Spohn,2026Spohn,2026UmbilicMultilane}.   

Here we obtain a set of sufficient conditions for the rates of a multilane ASEP to have a time-stationary Bernouilly  measure,  Eq.(\ref{RatesBernoulliMultilane}) and calculate 
the respective stationary current,  Eq. (\ref{SteadyCurrentMultilane}).  The expression for the stationary current (\ref{SteadyCurrentMultilane})
has appeared,  without  a proof,  in  \cite{2015Fibonacci}.  However,  it was interpreted within a setup of 
uni-directional particle hoppings,  which restricted the allowed range of interaction constants  to a part of the real axis.  
In present communication  we bridge this gap by proving the  formula (\ref{SteadyCurrentMultilane}).  In addition,  we show 
 that  (\ref{SteadyCurrentMultilane}) has a stochastic interpretation (the underlying microscopic hopping rates) for  interaction constants  of unrestricted range,  taking the values over the whole real axis.

\begin{figure}[tbp]
\centerline{
\includegraphics[width=0.5\textwidth]{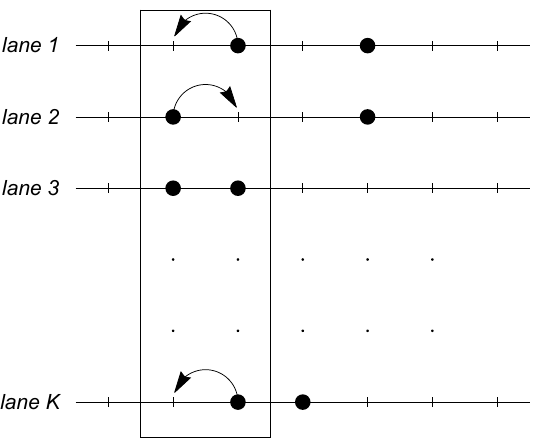}
}
\caption{ Schematic setup of a Multilane ASEP model.   Each particle hops to nearest neighbouring sites on the same 
lane if it is empty.  Rates of selected hoppings,  shown by arrows,  depend on the configuration of particles within the  rectangle on all the lanes.  For Bernoullli measure to be stationary the rates should satisfy the condition (\ref{RatesBernoulliMultilane}),
see text.
}
\label{FigMultilanes}
\end{figure}

\textit{Multilane ASEP with Bernouilly stationary measure.~~}
Let us consider a multilane ASEP: a system of parallel chains (lanes),  where each chain contains hardcore particles hopping randomly to their nearest neighbouring sites on the same chain,  provided they are empty,  
with some rates,  see Fig.~\ref{FigMultilanes}.     
Hoppings between the chains are forbidden as well
as occupation of a site by more than one particle (exclusion principle).  
Let us assume,  then,  
  that stationary probabilities $P(C)$ of  all  particle configurations
are equal,

\begin{align}
&P(C)=P(C'), \quad \forall C,C'. \label{Bernoulli}
\end{align}
The steady state  probabilities $ P(C)$ of a Markov process satisfy the  well-known stationary Master equation
\begin{align}
&\sum_{C'\neq C} r_{C',C} P(C') = P(C) \sum_{C''\neq C} r_{C,C''}   \label{SteadyW}
\end{align}
or,  using (\ref{Bernoulli}),
\begin{align}
&\sum_{C'\neq C} r_{C',C} =  \sum_{C''\neq C} r_{C,C''},   \label{SteadyBernoulli}
\end{align}
where $C$ is arbitrary configuration,  $C'$ are configurations from which one can reach $C$ via elementary hopping
with rate $r_{C',C} $ 
and $C''$ are configurations reachable from $C$,   with rates $r_{C,C''}$.

\begin{figure}[tbp]
\centerline{
\includegraphics[width=0.5\textwidth]{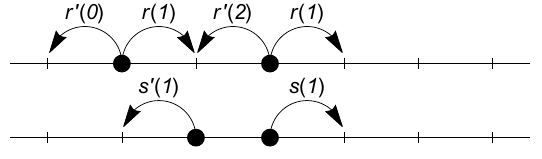}
}
\caption{ Schematic setup of a two-lane ASEP model with hopping rates $r(d),r'(d)$ and $s(d),s'(d)$, see text.
}
\label{Fig2lanes}
\end{figure}

To obtain the conditions for Bernoulli stationary measure in the  multilane ASEP it is enough to consider, first,  just two interacting lanes,  parallel to each other, see  Fig. ~\ref{Fig2lanes}.
Denote by $n_k$ and $m_k$ particle occupation numbers on site $k$ of lane $1$ and lane  $2$ respectively.  
We adopt an exclusion rule: each lattice site can either be empty or be occupied by  one particle, 
$n_k=0,1$ and $m_k=0,1$.  
The hopping rates are postulated as follows: 
 allowable hoppings concern nearest neighbouring sites at the same lane
(a lane change is not allowed),
with exclusion rule: if site $k$ is occupied ($n_k=1$),  a particle at site $k$ on lane $1$ can hop to the right to the site $k+1$  if it is empty ($n_{k+1} =0$),
with rate $r_{C,C'} \equiv r(m_k+m_{k+1})$,  which depends just on sum of occupation numbers on the other lane $m_k+m_{k+1}=0,1,2$.   
A particle hops left from site $k+1$ to site $k$  with rate $r'(m_k+m_{k+1})$.  
 
Likewise,  a particle on lane $2$ performs nearest neighbouring hoppings from site $k\rightarrow k+1$
and $k+1\rightarrow k$ with rates  $s(n_k+n_{k+1})$
and $s'(n_k+n_{k+1})$ respectively.  The two-lane ASEP
setup with corresponding rates is illustrated in Fig.~\ref{Fig2lanes}.

For the following we adopt the following notation: 
\begin{align}
&\tilde r(d) = r (d)-r'(d),  \no \\
&\tilde s(d) = s (d)-s'(d).  \label{rtilde}
\end{align}
Note that while $r(d),r'(d),s(d),s'(d)$ are,  by definition of hopping rate, nonnegative,  $\tilde r(d)$ and 
$\tilde s(d)$ are arbitrary real numbers.

Let us consider e.g.  a configuration $C$ with $3$ particles only  $n_{k+1}=n_{k}=m_{k}=1$,  and write down 
Eq (\ref{SteadyBernoulli}).  We obtain   
\begin{align}
&\tilde r(1) - \tilde r(0)=\tilde s(2)-\tilde s(1). \label{eq1}
\end{align}
For any other configuration $C$ we generically find that a condition 
 \begin{align}
&\tilde r(d+1) - \tilde r(d)=\tilde s(c+1)-\tilde s(c) \label{eq2} 
\end{align}
must be satisfied, for any $c,d$ taking values $0,1$.  An obvious solution of the above is to set

\begin{align}
&\tilde r(d) = A+ \frac{a}{2} \, d,  \label{RatesBernoulli} \\
&\tilde s(d) =B +  \frac{a}{2} \, d, \no
\end{align}
where $A,B,a$ are arbitrary real constants, and the coefficient $\frac12$ is introduced for a  convenience. 
Note that indeed $A= \tilde r(0) = r(0) -r'(0)$,  $B=s(0)-s'(0)$,   $a= r(2)-r'(2) - A $ can have 
any sign, positive or negative.  The corresponding steady currents  $j_\al$ of lane $\al$ in the infinite system are easy to obtain.
Namely,  denoting average particle densities  on lanes $1,2$ as $\langle n_k \rangle=\rho_1$,      $\langle m_k \rangle=\rho_2$,
we have:
\begin{align}
&j_1 = \langle n_k(1-n_{k+1}) \rangle  \sum_{m_k,m_{k+1}=0}^1 r(m_k+m_{k+1}) \langle m_k m_{k+1} \rangle \no \\
& -  \langle n_{k+1}(1-n_{k}) \rangle  \sum_{m_k,m_{k+1}=0}^1  r'(m_k+m_{k+1}) \langle m_k m_{k+1} \rangle \no\\
&= \rho_1 (1-\rho_1)   \sum_{m_k,m_{k+1}=0}^1  \tilde r(m_k+m_{k+1}) \langle m_k m_{k+1} \rangle   \label{currentTilde} \\
&=\rho_1 (1-\rho_1) \left(   
A  (1-\rho_2)^2 + 2\left(A+ \frac{a}{2} \right)  \rho_2 (1-\rho_2) +(A+a)  \rho_2^2 
\right) \no \\
&= \rho_1 (1-\rho_1) \left(   
A+  a \rho_2 \right) \label{j1:final}
\end{align}
In the above we used absence of correlations in the Bernouilly stationary measure
$ \langle n_k (1-n_{k+1}) \rangle =  \langle n_k \rangle \langle 1-n_{k+1} \rangle=\rho_1(1-\rho_1)$,
$ \langle m_k m_{k+1} \rangle =  \langle m_k \rangle \langle m_{k+1} \rangle$,  leading e.g.  to $ \langle 0 1 \rangle=  \langle 1 0 \rangle=
\rho_2(1-\rho_2)$ etc.. 
Proceeding analogously for the second lane, we obtain
\begin{align}
&j_1(\rho_1,\rho_2) = \rho_1 (1-\rho_1) (A +a \rho_2), \no\\
&j_2(\rho_1,\rho_2) = \rho_2 (1-\rho_2) (B +a \rho_1),\no
\end{align}
where  the range of $A,B, a$ is the whole real axis, and $\rho_\al$ is the particle density on lane $\al$.

 Addition of further lanes to the above scheme is straightforward since the contributions to the 
Bernoulli steady state condition (\ref{SteadyBernoulli}) from further lanes results  simply in addition of further terms 
leading to additional  conditions of type (\ref{eq2}).    For each pair of lanes the interlane interaction constants $a$ can be 
chosen differently.     We shall denote the interaction constant between the lanes $\al,\mu$ as $\ga_{\al \mu} $. 
Consequently,  the Bernouilly stationary measure condition for the hopping rates of a multilane ASEP has the following form:
the difference $\tilde r_\al =  r_\al-r'_\al$  between the forward hopping rate $k \rightarrow k+1$ and the backward hopping rate $k \leftarrow k+1$ on lane $\al$ is given by
\begin{align}
&\tilde r_\al = B_\al +\sum_{\mu \neq \al}^K \frac{\ga_{\al \mu}}{2} (n^{\mu}_k + n^{\mu}_{k+1} ),  \label{RatesBernoulliMultilane}\\
&\ga_{\al \mu} = \ga_{\mu \al},\no
\end{align}
where  $B_\al$ is an arbitrary constant,   $n^{\mu}_k$ is a particle occupation number on site $k$ of lane $\mu$,
 and  parameters $\ga_{\al \mu} = \ga_{\mu \al}$  measure interaction between the lanes $\al,\mu$.    Note that $\ga_{\al \mu}$ can be positive or negative,  
and  $\ga_{\al \nu}=0$ means that mutual interaction between the lanes $\al,\nu$ is absent. 
This leads to the steady current on lane $\al$
 \begin{align}
&j_\al = \rho_\al (1-\rho_\al) \left(B_\al +\sum_{\mu \neq \al}^K  \ga_{\al \mu} \rho_\mu \right), \label{SteadyCurrentMultilane}\\
& \ga_{\al \mu}= \ga_{\mu\al},\no
\end{align}
 where both $B_\al $ and  $\ga_{\mu \al}$  lie on real axis.
To prove (\ref{SteadyCurrentMultilane}) we denote 
 \begin{align}
&I(A,a, \mu)= \sum_{n^\mu_k,n^\mu_{k+1}=0}^1  \left( A + \frac{a}{2} (n^\mu_k + n^\mu_{k+1})\right)  \langle n^\mu_k n^\mu_{k+1} \rangle.
\end{align}
We readily find,  see last passage leading to (\ref{j1:final}), that 
 \begin{align}
&I(A,a, \mu)= A + a \rho_\mu. \label{Iformula}
\end{align}
Steady current is given, analogously to (\ref{currentTilde}),  by
\begin{align}
&j_\al=\rho_\al (1-\rho_\al) Z,\no\\
& Z = 
 \sum_{n^{\mu_1}_k,n^{\mu_1}_{k+1}}   \sum_{n^{\mu_2}_k,n^{\mu_2}_{k+1}} 
 \cdots \left(
B_\al +   \frac{ \ga_{\al \mu_1}}{2} ( n^{\mu_1}_{k}+  n^{\mu_1}_{k+1}) +  \sum_{\mu\neq \al,\mu_1}\frac{\ga_{\al \mu} }{2}( n^{\mu}_{k}+  n^{\mu}_{k+1})   \right) 
\langle n^{\mu_1}_k  n^{\mu_1}_{k+1} \rangle
\langle n^{\mu_2}_k  n^{\mu_2}_{k+1} \rangle \cdots \no
\end{align}
Performing the first summation  and using (\ref{Iformula}) we obtain
\begin{align}
& Z=  \sum_{n^{\mu_2}_k,n^{\mu_2}_{k+1}}  \sum_{n^{\mu_3}_k,n^{\mu_3}_{k+1}} 
 \cdots \left( B_\al +  \ga_{\al \mu_1} \rho_{\mu_1}+ \sum_{\mu\neq \al,\mu_1}\frac{\ga_{\al \mu}}{2} ( n^{\mu}_{k}+  n^{\mu}_{k+1})  \right)
\langle n^{\mu_2}_k  n^{\mu_2}_{k+1} \rangle  \langle n^{\mu_3}_k  n^{\mu_3}_{k+1} \rangle\cdots
\end{align}
Iterating the above procedure  we obtain $Z=B_\al +\sum_{\mu\neq \al}  \ga_{\al, \mu} \rho_\mu$ leading to   Eq. (\ref{SteadyCurrentMultilane}).

\begin{acknowledgments}
 V.P.  acknowledges support by ERC Advanced grant
 No.~101096208 -- QUEST,  Research Program P1-0402 and Grant N1-0368 of Slovenian Research and Innovation Agency (ARIS) and by Deutsche Forschungsgemeinschaft through DFG project KL645/20-2. 
\end{acknowledgments}


\begin{thebibliography}{99}
\bibitem{Pipkin} MacDonald J T, Gibbs J H and Pipkin A C  \textit{Biopolymers } \textbf{6} 1 (1968)

\bibitem {2006MallickASEP}
Olivier Golinelli and Kirone Mallick, The asymmetric simple exclusion process: an integrable model for non-equilibrium statistical mechanics,
 \textit{J. Phys. A: Math. Gen.  } \textbf{39}, 12679 ( 2006)

\bibitem{2011MallickASEP} 
 Kirone Mallick,  Some exact results for the exclusion process, 
 \textit{J. Stat. Mech. } P01024 (2011)




\bibitem{2000Johansson}
K. Johansson,  Shape fluctuation and random matrices, \textit{Comm. Math. Phys. } \textbf{209}, 437 (2000).


\bibitem{2001GunterReview} Sch\"utz G. M. 2001 Exactly solvable models for many-body systems far from 
equilibrium, in: Phase Transitions and Critical Phenomena. Vol. 19, eds C Domb and J
Lebowitz (London: Academic Press)
 
\bibitem {2004MultilanePopkovSalerno} V. Popkov and M. Salerno, Hydrodynamic limit of multi-chain driven
diffusive models, \textit{ Phys. Rev. E }\textbf{69}, 046103 (2004)


\bibitem{2015Fibonacci} V. Popkov, A. Schadschneider, J. Schmidt,  and G.M. Sch\"{u}tz,
Fibonacci family of dynamical universality classes,
 \textit{PNAS}, vol. \textbf{112}  no. 41, 12645-12650 (2015),



\bibitem{2024Spohn} 
Dipankar Roy, Abhishek Dhar, Konstantin Khanin, Manas Kulkarni and Herbert Spohn,
Universality in coupled stochastic Burgers systems with degenerate flux Jacobian,
\textit{J.Stat. Mech}  033209 (2024)

\bibitem{2026Spohn} 
Herbert Spohn
The Popkov-Schütz two-lane lattice gas: Universality for general jump rates
 \textit{J. Stat. Mech. } 023203 (2026).

\bibitem{2026UmbilicMultilane}
J. Schmidt,  Z. Krajnik and V. Popkov, 
Universality in driven systems with a
multiply-degenerate umbilic point,
\textit{J.Stat. Mech.  } 033202 (2026),



\end{thebibliography}
\end{document}